\documentclass[10pt,english,superscriptaddress,twocolumn]{revtex4}

\usepackage[T1]{fontenc}
\usepackage[latin9]{inputenc}
\usepackage{color}
\usepackage{babel}

\usepackage{amsmath}
\usepackage{graphicx}
\usepackage{amssymb}
\usepackage{esint}
\usepackage[unicode=true, pdfusetitle,
 bookmarks=true,bookmarksnumbered=false,bookmarksopen=false,
 breaklinks=false,pdfborder={0 0 0},backref=false,colorlinks=false]
 {hyperref}

\makeatletter

\providecommand{\tabularnewline}{\\}

\@ifundefined{textcolor}{}
{%
 \definecolor{BLACK}{gray}{0}
 \definecolor{WHITE}{gray}{1}
 \definecolor{RED}{rgb}{1,0,0}
 \definecolor{GREEN}{rgb}{0,1,0}
 \definecolor{BLUE}{rgb}{0,0,1}
 \definecolor{CYAN}{cmyk}{1,0,0,0}
 \definecolor{MAGENTA}{cmyk}{0,1,0,0}
 \definecolor{YELLOW}{cmyk}{0,0,1,0}
 }

\makeatother

\begin{document}

\title{Di-electron production from vector mesons with medium modifications
in heavy ion collisions}

\author{Hao-jie Xu}

\affiliation{Department of Modern Physics, University of Science and Technology
of China, Anhui 230026, People's Republic of China}

\author{Hong-fang Chen}

\affiliation{Department of Modern Physics, University of Science and Technology
of China, Anhui 230026, People's Republic of China}

\author{Xin Dong}

\affiliation{Nuclear Science Division, Lawrence Berkeley National Laboratory,
Berkeley, California 94720, USA}

\author{Qun Wang}

\affiliation{Department of Modern Physics, University of Science and Technology
of China, Anhui 230026, People's Republic of China}

\author{Yi-fei Zhang}

\affiliation{Department of Modern Physics, University of Science and Technology
of China, Anhui 230026, People's Republic of China}
\begin{abstract}
We reproduce the di-electron spectra in the low and intermediate mass
regions in most central Au+Au collisions by the STAR Collaboration
incorporation of the STAR detector acceptance. We also compare our
results with PHENIX data constrained by the PHENIX acceptance. We
include the medium modifications of vector mesons from scatterings
of vector mesons by mesons and baryons in the thermal medium. The
freezeout contributions from vector mesons are also taken into account.
The space-time evolution is described by a 2+1 dimensional ideal hydrodynamic
model. The backgrounds from semi-leptonic decays of charm hadrons
are simulated by the PYTHIA event generator and corrected by the nuclear
modification factor of electrons from charm decays. It is difficult
to extract the thermal contributions from those from charm decays
in the invariant mass spectra alone and in the current detector acceptances.
Other observables such as transverse momenta and collective flows
may provide additional tools to tag these sources. 
\end{abstract}
\maketitle

\section{Introduction}

The electromagnetic probes such as photons and dileptons are expected
to provide clean signatures for the quark gluon plasma (QGP) in heavy
ion collisions due to their instant emissions once produced \cite{McLerran:1984ay,Kajantie:1986dh,Rapp:1999ej,Alam:1999sc,vanHees:2007th,Chatterjee:2007xk,Dusling:2008xj}.
These thermal photons and dileptons contain undistorted information
about the space-time trace of the new state of matter formed in such
collisions. The invariant mass spectrum is usually divided into the
low, intermediate and high mass regions (LMR, IMR and HMR), based
on the notion that each region is dominated by different sources of
dileptons. In the LMR, $M\lesssim$1 GeV, dileptons are mainly from
vector meson decays and may be related to chiral symmetry restoration
\cite{Pisarski:1981mq,Martell:2004gt,Arnaldi:2006jq,vanHees:2006ng,Ruppert:2007cr,Ghosh:2010wt}.
In the HMR, $M\gtrsim3$ GeV, dileptons are dominated by the Drell-Yan
process and quarkonium decays. In the IMR, $1\lesssim M\lesssim3$
GeV, it was argued that dileptons from semi-leptonic decays of correlated
open charm hadrons are dominant \cite{:2008asa}.

The medium modifications of the $\rho$ meson spectral functions are
successful in describing the di-muon enhancement in the LMR of the
NA60 experiment at the SPS energy \cite{vanHees:2006ng,Ruppert:2007cr,Dusling:2006yv}.
The PHENIX and STAR collaborations also observed such an enhancement
in the di-electron spectra at the RHIC energy \cite{Adare:2009qk,Zhao:2011wa}.
The thermal quark-antiquark annihilation in the QGP phase is expected
to give a measurable signal in the IMR for the deconfinement phase
transition at RHIC energy \cite{Deng:2010pq}. However, in this mass
region, the di-lepton yield from semi-leptonic decays of open charm
mesons increases rapidly with the collisional energy. The single leptons
from open charm mesons and their dynamic correlations are expected
to undergo medium modifications. The question is: to what extent the
di-leptons from charm hadrons with medium modifications mix up with
the thermal contributions from the QGP in the IMR. Another issue is
that the dilepton spectra measured by the STAR and PHENIX collaborations
are very different in the LMR. It is worthwhile to to look at this
disagreement closely by using the Monte Carlo simulation incorporating
the different acceptances of STAR and PHENIX detectors. 

In this paper, we try to reproduce the data of di-electron invariant
mass spectra in the LMR and IMR in central Au+Au collisions at 200
GeV. We will include the medium modifications of the vector mesons
and charm hadrons. The acceptances of STAR and PHENIX detectors are
incorporated in our calculation. We will use a 2+1 dimension ideal
hydrodynamic model to give the space-time evolution of the fireball,
where the parameters are determined by fitting the data of transverse
momenta of long life hadrons (pions, Kaons and protons). The spectra
of charm hadrons ($D^{0}$,$D^{\pm}$, $D_{s}$ and $\Lambda_{c}$)
are given by a simulation of the PYTHIA event generator. We neglect
the Dalitz decay channel for $\pi^{0}$: $\pi^{0}\rightarrow e^{+}e^{-}\gamma$
but include those for $\eta$ and $\omega$: $\eta\rightarrow e^{+}e^{-}\gamma$
and $\omega\rightarrow e^{+}e^{-}\pi^{0}$. The contribution from
pion's Dalitz decay is mainly below $m_{\pi}$ and irrelevant to our
current range of the invariant mass.

\section{Parameters in hydrodynamic model }

\label{sec:Hydrodynamic}We use a 2+1 dimensional ideal hydrodynamic
model \cite{Deng:2010pq} to give the space-time evolution of the
medium created in heavy ion collisions. We choose two types of the
Equation of state (EOS) \cite{Hirano:2002ds,Huovinen:2009yb,Bazavov:2009zn,Shen:2010uy},
S95P-CE (CE) with complete chemical equilibrium to very low temperatures
and a wide range of phase transition temperatures from 184MeV to 220MeV,
and S95P-PCE (PCE) with partial chemical equilibrium below chemical
freezeout temperature $T_{chem}=165$ MeV %
\footnote{The EOS tables (by P. Huovinen), their analytic parametrizations (by
T. Riley and C. Shen) and the list of included hadrons are available
at the URL {[}https://wiki.bnl.gov/hhic/index.php/Lattice\_calculatons
\_of\_Equation\_of\_State{]}. %
}. After kinetic freeze-out, we use the Cooper-Frye formula \cite{Cooper:1974mv,Kolb:2000sd}
to obtain the momentum spectra for each hadron species \begin{equation}
E\frac{dN_{i}}{d^{3}p}=\frac{dN_{i}}{dyp_{T}dp_{T}d\phi}=\frac{g_{i}}{(2\pi)^{3}}\int_{T_{f}}d\Sigma_{\mu}p^{\mu}n_{i}(x,u\cdot p),\end{equation}
where $\Sigma_{\mu}$ denotes the normal vector of the freezeout hypersurface,
$T_{f}$ is the kinetic freeze-out temperature on the freezeout hypersurface,
$n_{i}$ is the phase space distribution function for the baryon/meson
species $i$ which can be Fermi-Dirac/Bose-Einstein distribution and
$g_{i}$ is its degeneracy factor, and $p_{T}$ denotes the transverse
momentum. 

Beside the EOS, there are some free parameters which should be fixed
in the hydrodynamical model, such as equilibration time $\tau_{0}$,
the initial energy density $e_{0}$ (or initial temperature $T_{0}$)
and the kinetic freeze-out energy density $e_{f}$ (or $T_{f}$).
At the RHIC energy $\sqrt{s_{NN}}=200$ GeV for Au+Au collisions \cite{:2008ez,Adler:2003au},
we constrain these parameters with STAR and PHENIX data for the rapidity
densities of multiplicities, $dN_{i}/dy$, and the $p_{T}$ spectra
for long-life hadrons (pions, koans and protons). The distribution
of the initial energy density is determined by the Glauber model with
5\% of the contribution from binary collisions. We focus on most central
collisions with impact parameter $b=2.4$ fm throughout the paper
and assume a system with vanishing net baryon number. 

\begin{figure}
\includegraphics[scale=0.4]{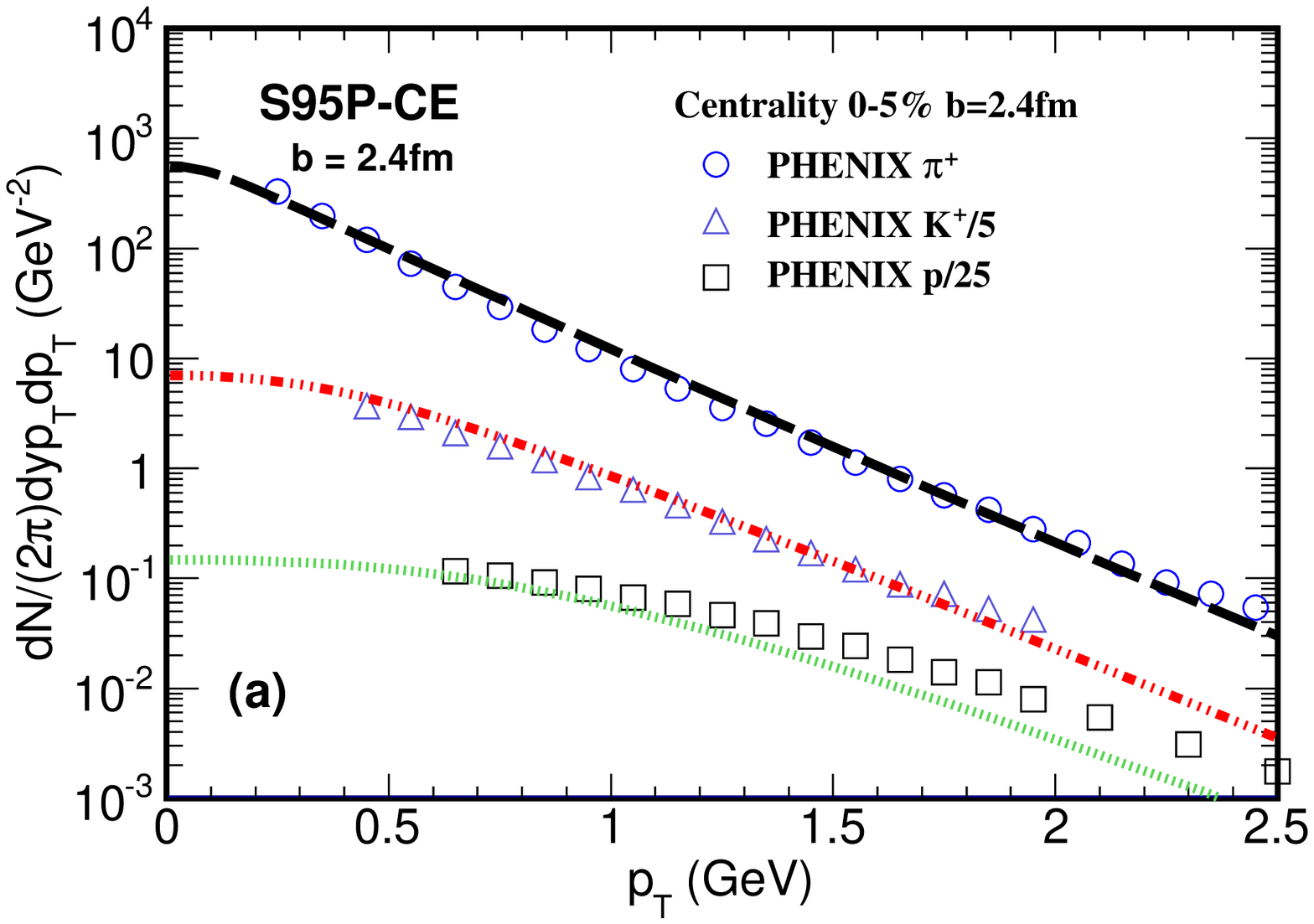}

\includegraphics[scale=0.4]{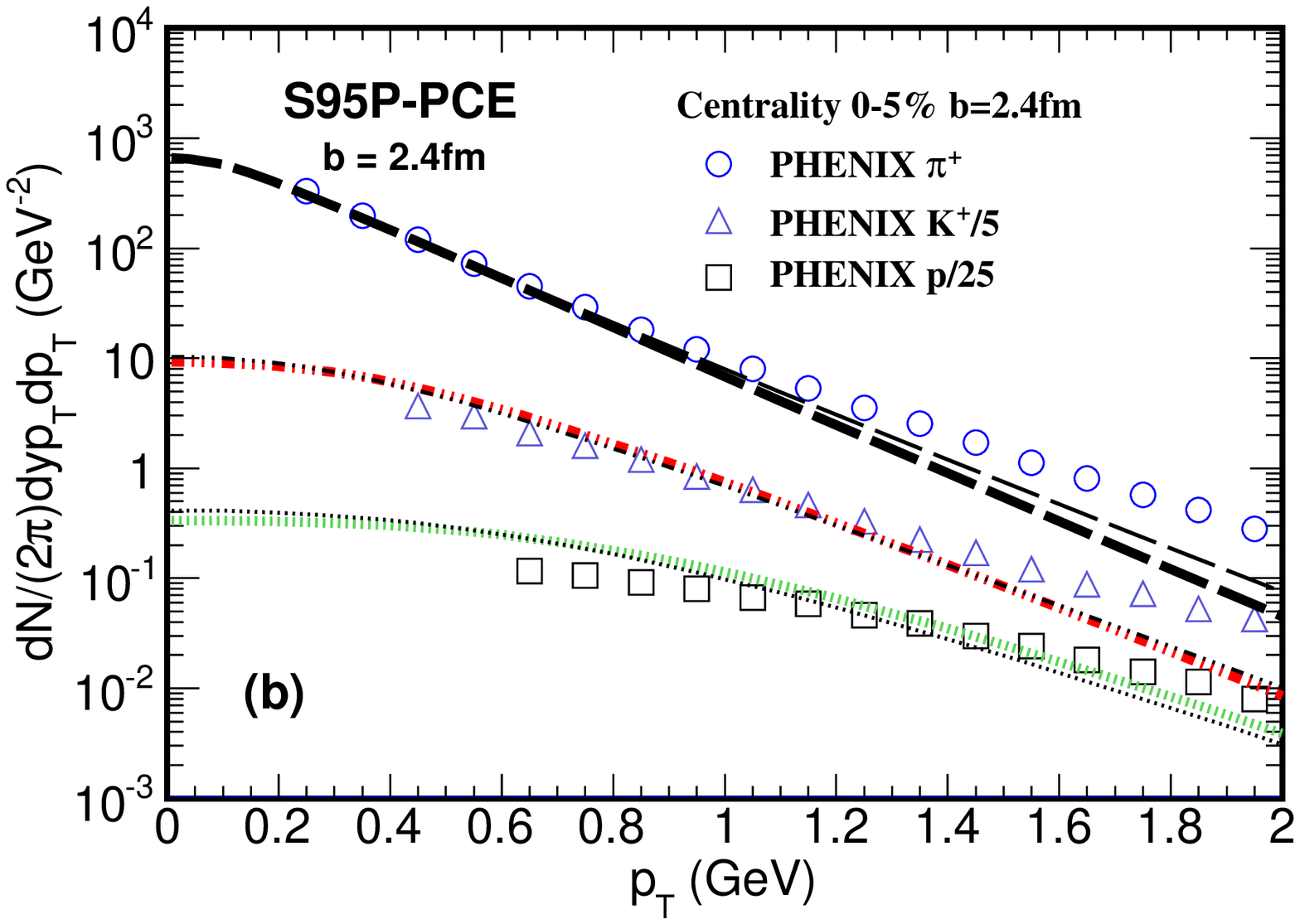}\caption{(Color online) Transverse momentum spectra for $\pi^{+}$ (black-dashed),
$K^{+}$ (red-dash-dot-dot-dotted) and $p$ (green-dotted) for central
collisions with centrality 0-5\%. The data are from the PHENIX collaboration
\cite{Adler:2003au}. (a) S95P-CE and (b) S95P-PCE EOS are used. In
(b), the differences between $T_{f}=106$ MeV (thick lines) and $T_{f}=$136
MeV (thin lines) in the low $p_{T}$ range are small.\label{fig:HadronSpectra}}

\end{figure}

To make a comparison, we choose the same initial conditions for two
EOS ($\tau_{0}=0.4$ fm , $e_{0}=45$ GeV/$\mathrm{fm^{3}}$ (or $T_{0}$=395
MeV)), but freeze-out conditions are different due to different relations
between the energy density and temperature. The same initial conditions
imply that the entropy densities are chosen to be the same for these
two EOS. We choose $T_{f}=136$ MeV ($e_{f}=0.12$ GeV/$\mathrm{fm^{3}}$)
for CE and roughly reproduce the $p_{T}$ spectra for the pions and
kaons, as shown in Fig. (\ref{fig:HadronSpectra}(a)). But the proton
yield is under-estimated. For PCE, we choose the kinetic freeze-out
occurs at either the same energy density or the same temperature as
CE, i.e. $(e_{f},T_{f})=$ ($0.12$ $\mathrm{GeV/fm^{3}}$, 106 MeV)
or ($0.275$ $\mathrm{GeV/fm^{3}}$, 136 MeV). The $p_{T}$ spectra
with PCE are shown in Fig. (\ref{fig:HadronSpectra}(b)). The parameter
sets are listed in Tab. (\ref{tab:Parameter-sets}).

The PCE scenario has a feature compared to the CE one \cite{Hirano:2002ds}:
$dN_{i}/dy$ and $p_{T}$ spectra in the low $p_{T}$ range ($p_{T}<1$
GeV) are almost independent of $T_{f}$. However, the high $T_{f}$
will reduce the dilepton production rate by shortening the evolution
time. So we can regard $T_{f}$ as a tuning parameter to the dilepton
production rate from the in-medium vector meson decay. 

\begin{table}
\centering{}\begin{tabular}{|c|c|c|c|c|}
\hline 
set & EoS & $T_{chem}[\mathrm{MeV}]$ & $T_{f}\left[\mathrm{MeV}\right]$ & $e_{f}[\mathrm{GeV/fm^{3}}]$\tabularnewline
\hline
\hline 
S1 & S95P-CE & - & 136 & 0.12\tabularnewline
\hline 
S2 & S95P-PCE & 165 & 136 & 0.275\tabularnewline
\hline 
S3 & S95P-PCE & 165 & 106 & 0.12\tabularnewline
\hline
\end{tabular}\caption{Parameter sets. We choose most central collisions with $b=2.4$ fm.
The initial conditions are chosen to be: $\tau_{0}=0.4$ fm , $e_{0}=45$
GeV/$\mathrm{fm^{3}}$ ($T_{0}$=395 MeV). \label{tab:Parameter-sets}}

\end{table}

We give a few comments about the yields of vector mesons, whose multiplicity
ratios, $\rho/\pi$,$\omega/\pi$ and $\phi/\pi$, are given in Tab.
(\ref{tab:mulitiplicity}). Most of the decays of the $\omega$ and
$\phi$ mesons take place after the kinetic freeze-out. The yields
of $\omega$ and $\phi$ are in good agreement with the data and will
contribute to the dilepton rate. We see that the ratio $\rho/\pi$
at the freeze-out is smaller than the data, but we note that most
of the $\rho$ mesons decay in the thermal medium earlier than the
kinetic freeze-out whose contribution to the dilepton production dominates
the dilepton spectra. Hereafter we will use the parameter set S3 of
PCE as a default choice unless stated explictly. The comparison will
be made with other parameter sets. 

\begin{table}
\centering{}\begin{tabular}{|c|c|c|c|c|c|}
\hline 
sets & $\pi^{+}$ & $p$ & $\rho/\pi$  & $\omega/\pi$  & $\phi/\pi$\tabularnewline
\hline
\hline 
S1  & $285.7$ & $12.2$ & $7.91\times10^{-2}$  & $7.49\times10^{-2}$  & $1.86\times10^{-2}$\tabularnewline
\hline 
S2 & $270.2$ & $23.6$ & $7.78\times10^{-2}$  & $9.85\times10^{-2}$  & $2.89\times10^{-2}$\tabularnewline
\hline 
S3 & $261.7$ & $24.6$ & $5.50\times10^{-2}$ & $9.82\times10^{-2}$  & $3.13\times10^{-2}$\tabularnewline
\hline 
PHENIX  & $281.8$  & $18.4$ & $1.03\times10^{-1}$  & $8.98\times10^{-2}$  & $2.14\times10^{-2}$\tabularnewline
\hline 
STAR  & $327$  & $34.7$ & $1.69\times10^{-1}$  & -  & $2.65\times10^{-2}$\tabularnewline
\hline
\end{tabular}\caption{Rapidity densities $dN_{i}/dy$ for meson/proton yields in most central
collisions with $b=2.4$ fm. The PHENIX data are taken from Ref. \cite{Adler:2003au,Adare:2009qk},
where they only have the $\pi^{+}$ data in most central collisions
and their $\rho$ data are from the fragmentation model. The STAR
data are from Ref. \cite{Adams:2003cc,:2008fd}.\label{tab:mulitiplicity}}

\end{table}

\section{Dilepton Emissions in Heavy Ion Collisions}

In the thermalized medium, hadron gas (HG) or quark gluon plasma (QGP),
the dilepton production rate per unit volume is given by \begin{align}
\frac{dN_{ll}}{d^{4}xd^{4}p} & =-\frac{\alpha}{4\pi^{4}}\frac{1}{M^{2}}n_{B}(p\cdot u)\left(1+\frac{2m_{l}^{2}}{M^{2}}\right)\nonumber \\
 & \times\sqrt{1-\frac{4m_{l}^{2}}{M^{2}}}\mathrm{Im}\Pi^{R}\left(p,T\right).\end{align}
 Here $m_{l}$ is the lepton mass, $\alpha=e^{2}/4\pi$ is the fine
structure constant with the electric charge $e$ for leptons, $p=(p_{0},\mathbf{p})=p_{1}+p_{2}$
is the dilepton 4-momentum and $M=\sqrt{p^{2}}$, $n_{B}=1/\left(e^{p\cdot u/T}-1\right)$
($T$ and $u$ are the local temperature and fluid velocity respectively)
is the Bose distribution function, $\Pi_{\mu\nu}^{R}$ is the retarded
photon polarization tensor from the quark or hadronic loop, and $\Pi^{R}=\frac{1}{3}\Pi_{\mu}^{R\mu}$
. For the partonic phase, $\Pi^{R}$ given by the Born term reflects
the lowest order process $q\bar{q}\rightarrow\gamma^{*}\rightarrow l^{+}l^{-}$.
For the hadronic phase, $\Pi^{R}$ is further related to the retarded
vector-meson propagator $D_{V}^{R}$ with $V=\rho,\omega,\phi$ via
$\mathrm{Im}\Pi^{R}=-\left(e^{2}m_{V}^{4}/g_{V}^{2}\right)\mathrm{Im}D_{V}^{R}$
, where $g_{V}$ is the photon-vector-meson coupling constant in the
vector meson dominance model, and $m_{V}$ is the vector-meson mass.
The retarded vector meson propagator is \begin{equation}
\mathrm{Im}D_{V}^{R}=\frac{\mathrm{Im}\Pi_{V}^{R}}{\left(p^{2}-m_{V}^{2}+\mathrm{Re}\Pi_{V}^{R}\right)^{2}+\left(\mathrm{Im}\Pi_{V}^{R}\right)^{2}},\end{equation}
 where $\Pi_{R}$ is the contraction of the retarded vector meson
polarization tensor. 

\begin{figure}
\includegraphics[scale=0.4]{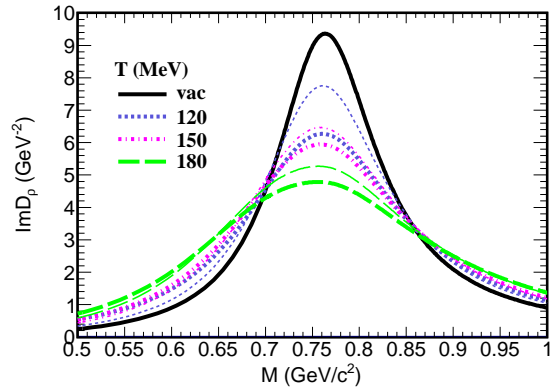}

\caption{The imaginary parts of the in-medium $\rho$ meson propagators (or
in-medium spectral functions) with (thick lines) and without (thin
lines) baryonic contributions. The chemical potentials in PCE EOS
are used. \label{fig:spectrafunction}}

\end{figure}

We use the hadronic many body effective theory \cite{Rapp:2000pe}
to calculate in-medium $\rho$ spectral functions by scattering with
surrounding mesons. Though we assume a net baryon free system at RHIC
energy, the effective chemical potentials in the PCE scenario will
give a considerable number of baryons \cite{Rapp:2002fc}, and we
assume that there are equal number of anti-baryons which give the
same contribution as the baryons to the $\rho$ spectral functions.\textcolor{red}{{}
}To include the baryonic (including anti-baryonic) contributions,
we use the empirical scattering amplitude method \cite{Eletsky2001}
which agrees with the hadronic many body effective theory \cite{Rapp2007}.
Here we only consider the coupling of the $\rho$ meson with baryonic
resonances in the medium and set the momentum $q=300$ MeV for the
$\rho$ meson in-medium propagator. The in-medium $\rho$ meson spectral
functions with and without baryonic contributions are shown in Fig.
(\ref{fig:spectrafunction}) at different temperatures but at a fixed
momentum $q=300$ MeV. The differences between with and without baryonic
contributions are larger at low temperatures than at high temperatures. 

Since the collision rate in a meson gas around the transition temperature
indicates a large broadening of $\phi$ meson spectra due to binary
collisions \cite{AlvarezRuso:2002ib}, we include this effect via
a schematic estimate as follows (with $T_{0}=150$ MeV), \begin{equation}
\Gamma_{\phi_{coll}}\simeq(22\:\mathrm{MeV})\left(\frac{T}{T_{0}}\right)^{6}.\end{equation}

\begin{figure}
\includegraphics[scale=0.4]{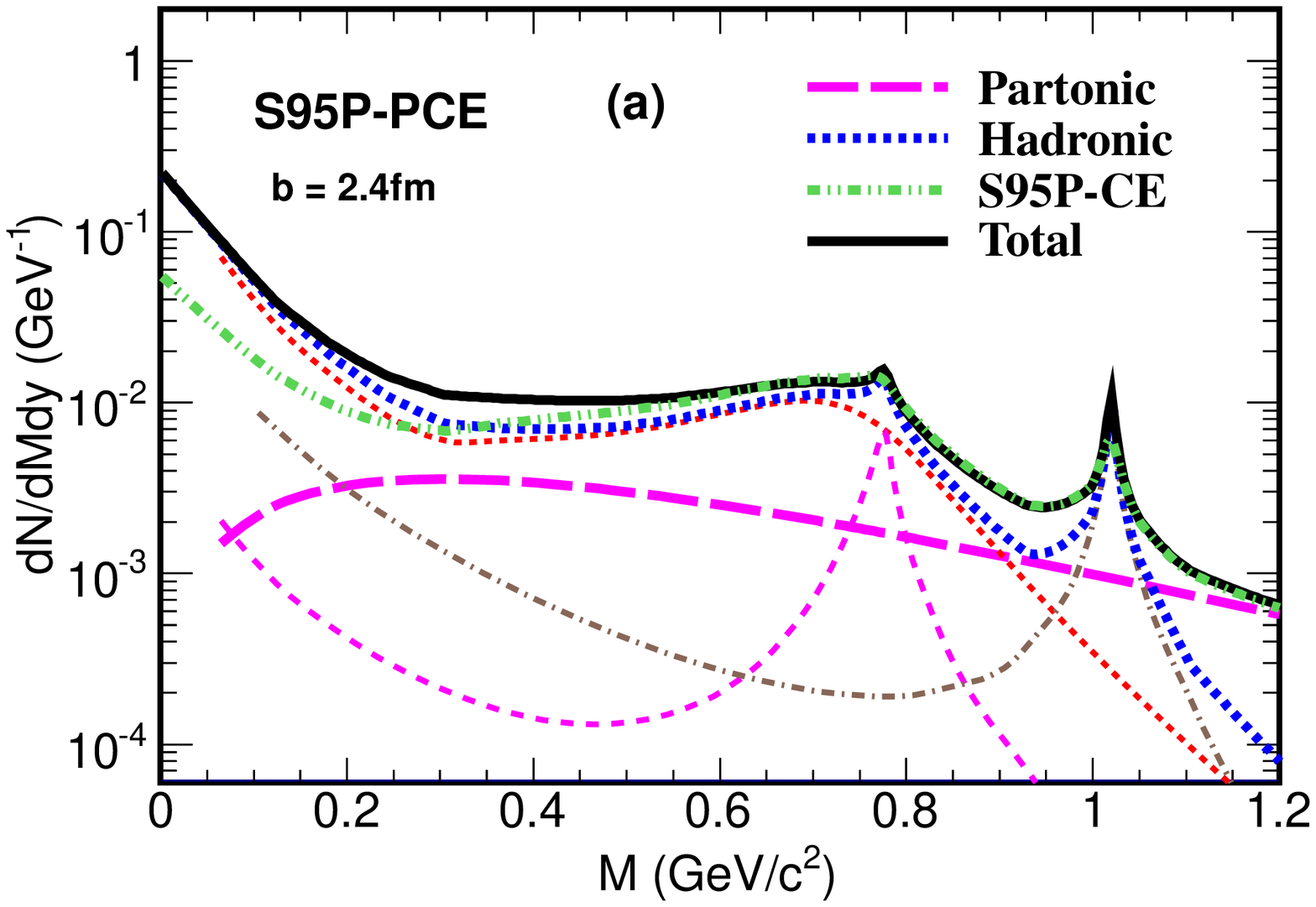}

\includegraphics[scale=0.4]{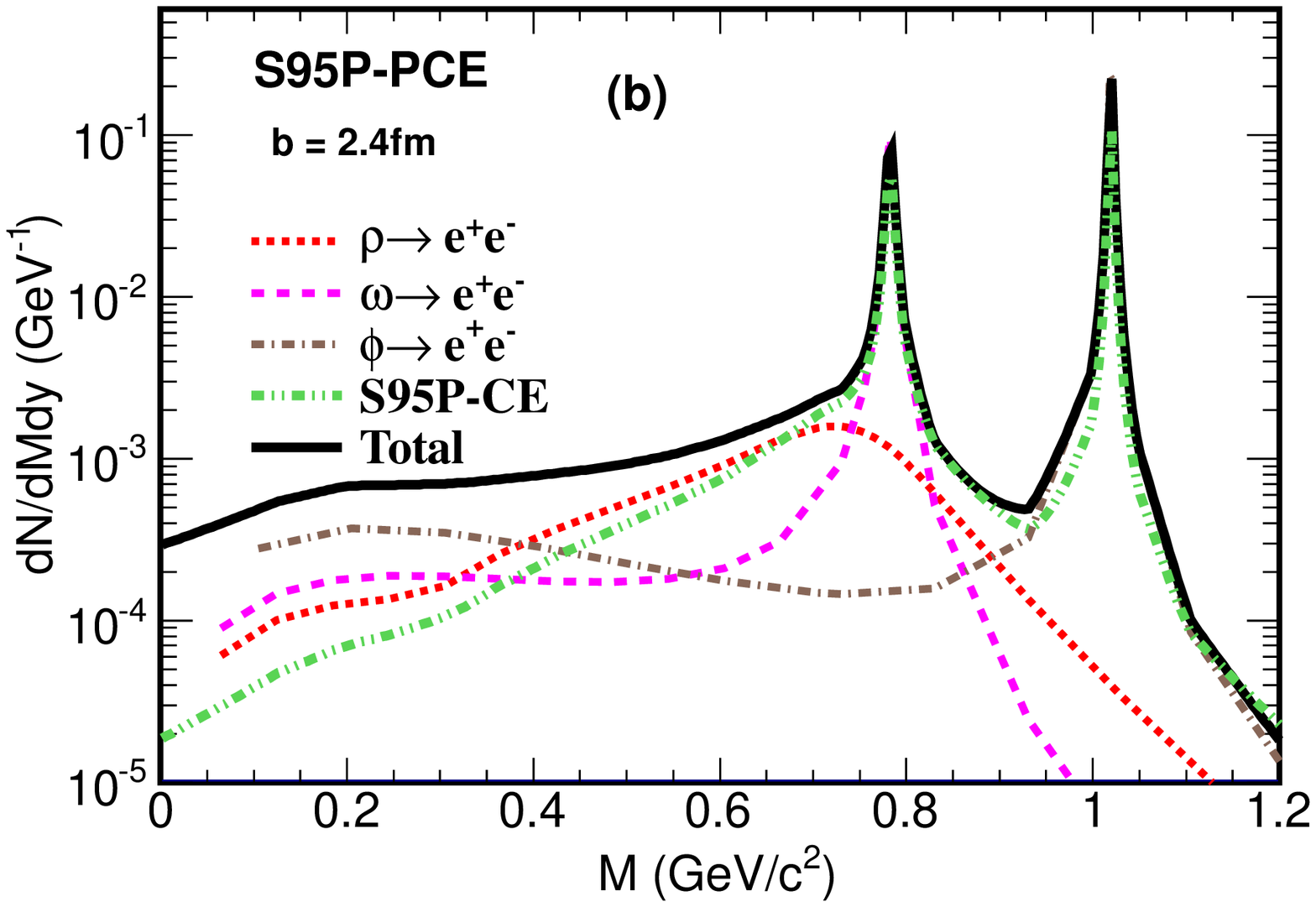}\caption{(Color online) (a) The invariant mass spectra of thermal di-electrons
in full phase space. In the partonic phase, the main source is $q\bar{q}\rightarrow\gamma^{*}\rightarrow e^{+}e^{-}$
(magenta-long-dashed line). In the hadronic phase, the total contribution
is the blue-dashed line, where the contribution from the $\rho$ meson
(red-dotted line) dominates, and those from the $\omega$ and $\phi$
mesons are shown in magenta-short-dashed and brown-dash-dotted lines
respectively. (b) The invariant mass spectra of di-electrons for vector
mesons $\rho$, $\omega$ and $\phi$ at the freezeout. \label{fig: EmissionRate}}

\end{figure}

In Fig. (\ref{fig: EmissionRate}(a)) we show the invariant mass spectra
of thermal di-electrons at the RHIC energy 200 GeV for most central
Au+Au collisions. Beside the $\rho$ component (red-dotted line) in
the hadronic phase in an early study \cite{Deng:2010pq}, we include
the in-medium $\omega$ (magenta-short-dashed line) and $\phi$ (light-brown-dash-dotted
line) contributions to thermal di-electrons. The thermal di-electrons
are dominated by the in-medium $\rho$ mesons, while the $\omega$
contribution is submerged under the broadened $\rho$ spectra.\textcolor{red}{{}
}The thermal spectra with the CE EOS (green-dash-dotted-dotted line)
has also been shown. The production rate in the PCE scenario is larger
in the invariant mass range below free $\rho$ mass than in the CE
one, though the temperatures with PCE are lower. This is because the
chemical potentials in the PCE scenario lead to a larger broadening
of the $\rho$ spectral function and an enhancement factor $e^{2\mu_{\pi}/T}$
compared with the CE scenario. The enhancement at low masses in the
PCE scenario is more obvious at lower $T_{f}$. We will come back
to this issue later. These two EOS have the same partonic contributions
because their differences only occur in the hadronic phase. 

The dilepton emission rate from the freezeout vector meson is given
by \begin{equation}
\frac{dN_{l\bar{l}}^{fo}}{d^{4}p}=\frac{\alpha}{3}\left(\frac{e}{g}\right)^{2}\frac{m_{V}}{\Gamma_{V}}\frac{dN_{V}^{fo}}{d^{4}p}.\end{equation}
where $\Gamma_{V}$ is the total decay width of the vector mesons.
The vector meson momentum spectra at thermal freezeout can be expressed
by the extended Cooper-Frye formula \cite{vanHees:2007th} \begin{equation}
\frac{dN_{V}^{fo}}{d^{4}p}=\frac{g_{s}^{\rho}}{4\pi^{4}}\int_{T_{f}}d\Sigma_{\mu}p^{\mu}\mathrm{Im}D_{V}n_{B}(p\cdot u).\end{equation}

Since the lifetimes of $\omega$ and $\phi$ are much longer than
the time scale of the freezeout process, we treat these contributions
as in vacuum and neglect the medium effect. The imaginary parts of
the $\omega$ and $\phi$ propagators can be given by the Breit-Wigner
formula, \begin{equation}
\mathrm{Im}D_{\omega,\phi}^{fo}=-\frac{m_{V}\Gamma_{V}}{(M^{2}-m_{V}^{2})^{2}+m_{V}^{2}\Gamma_{V}^{2}}.\end{equation}
But most of the $\rho$ mesons decay in the medium due to its short
lifetime, so we include the medium effect in the $\rho$ meson propagator.
In Fig.(\ref{fig: EmissionRate}(b)) are shown the invariant mass
spectra of di-electrons from the freezeout vector mesons, where the
sharp peaks of the $\omega$ and $\phi$ mesons can be seen compared
to a much broader bump of the $\rho$ meson.

\begin{figure}
\includegraphics[scale=0.4]{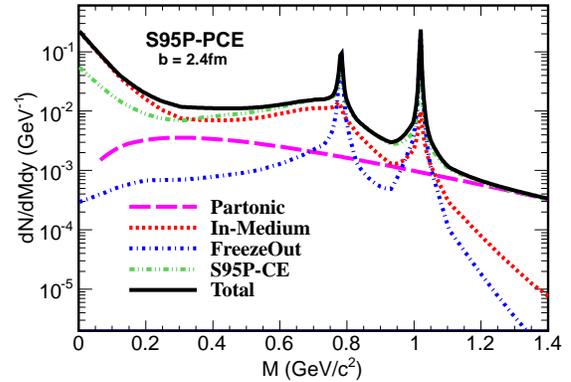}\caption{(Color online) The cocktail mixture of the partonic, in-medium and
freezeout hadronic sources for di-electrons. The partonic, in-medium
and freezeout hadronic contributions are in magenta-long-dashed, red-dotted,
and blue-dash-dotted lines, respectively. The total contribution is
in the black-solid line. \label{fig:invariatmass-total}}

\end{figure}

In Fig. (\ref{fig:invariatmass-total}) we sum over all sources we
have considered. The full mass spectra have two sharp peaks of the
$\omega$ and $\phi$ mesons at the freezeout same as in vacuum due
to their long lifetime. For comparison the in-medium spectra are also
shown where only a much lower peak from the $\phi$ meson is visible,
indicating clear medium effects. Subtracting these sharp peaks of
$\omega$ and $\phi$, the broadened spectrum of the in-medium $\rho$
meson can be seen. The partonic contribution dominates over the hadronic
one when $M>1.1$ GeV/$c^{2}$. These continuum-like IMR dileptons
may provide a direct probe to the deconfinement phase transition in
high energy heavy ion collisions. The different EOS give the similar
structure but slightly different magnitude. It seems that the low
mass enhancement favors the PCE scenario, we will come back to this
point with details later in the following section.

\section{Comparison with data}

Different from the SPS energy, the charm quarks have a considerable
production rate at the RHIC energy. So there is a large background
from semi-leptonic decays of the charm hadrons. In this section we
will estimate this background and compare our dilepton results with
the data. 

We use the event generator PYTHIA \cite{Sjostrand:2000wi} (version
6.416 with CTEQ5L PDF) to simulate the background from semi-leptonic
decays of the charm hadrons ($D^{0}$,$D^{\pm}$, $D_{s}$ and $\Lambda_{c}$).
The PHENIX collaboration also tuned the parameters of PYTHIA \cite{Adcox:2002cg}
to fit the charm hadron data at SPS and FNAL and single electron data
at ISR. The parameter dependences such as intrinsic $k_{T}$ and the
parton distribution functions are also addressed in Ref. \cite{:2008asa}.
In our paper, we do not consider the fluctuations from these parameter.

\begin{figure}
\includegraphics[scale=0.4]{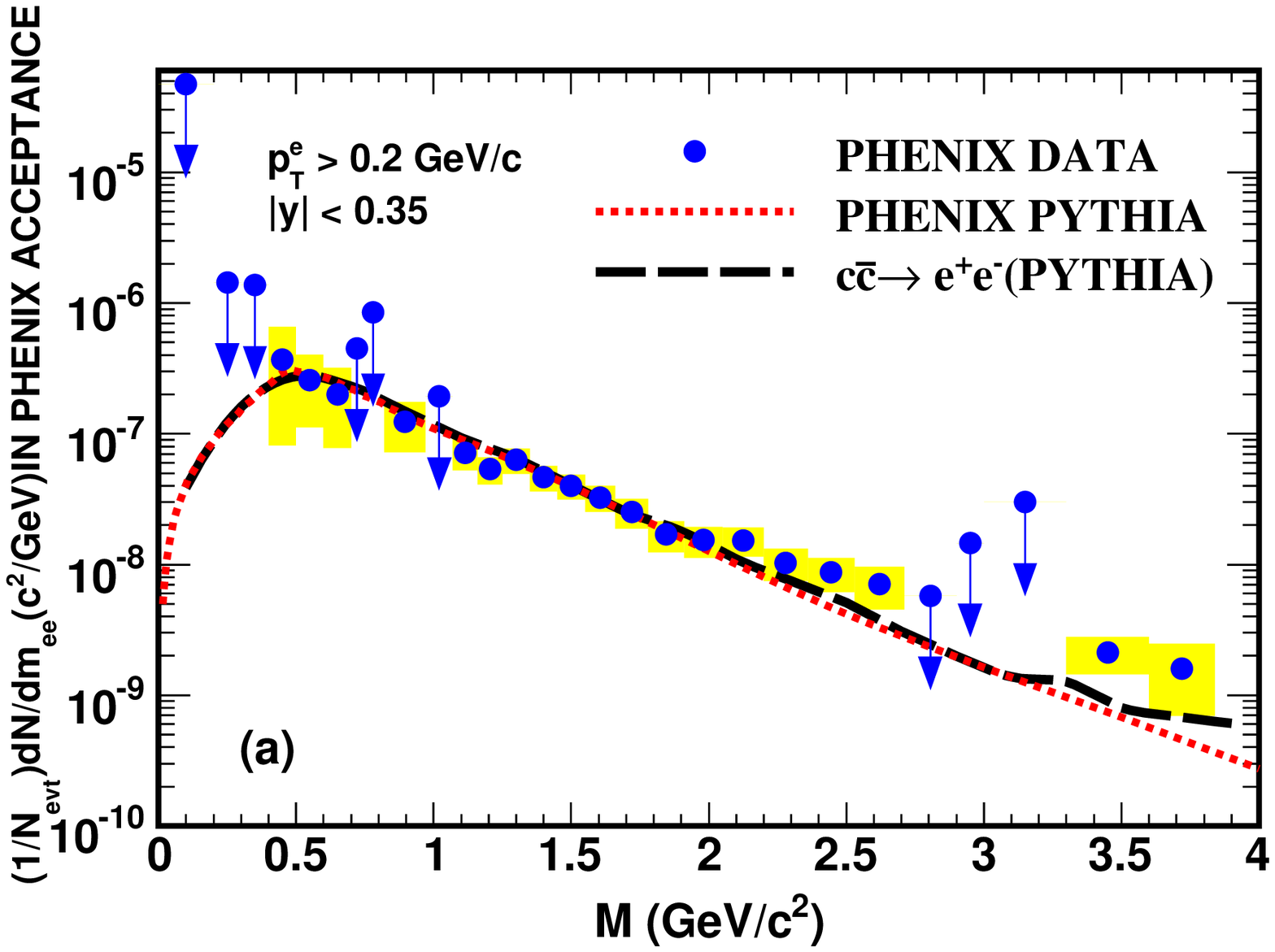}

\includegraphics[scale=0.4]{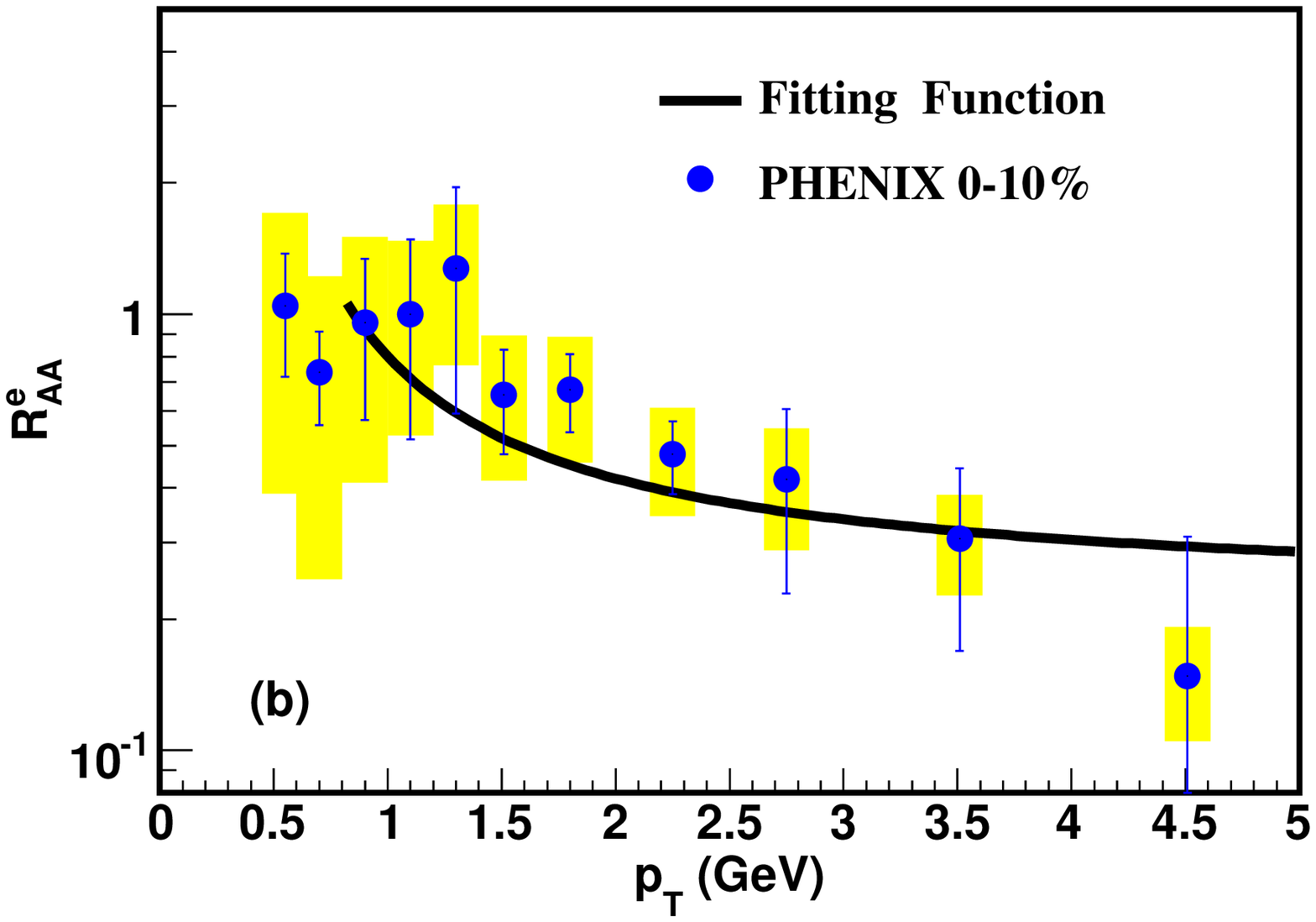} 

\caption{(Color online) The semi-leptonic decays of charm hadrons. (a) The
re-scaled di-electron cross section from charm hadrons of semi-leptonic
decays in p+p collisions by PYTHIA. The data are taken from the PHENIX
collaboration \cite{:2008asa}. (b) The nuclear modification factor
for nonphotonic electrons in central Au+Au collisions from the PHENIX
collaboration \cite{Abelev:2006db}. The fitting function is given
in Eq.(\ref{eq:raae}). \label{fig:invariantmass-charm}}

\end{figure}

In p+p collisions, the dilepton yield in the mass range {[}1.1,2.5{]}
GeV/$c^{2}$ is dominated by semi-leptonic decays of charm hadrons.
In the PHENIX acceptance the integrated yield of di-electrons per
event from heavy-flavor decays in that range is $(4.21\pm0.28\pm1.02)\times10^{-8}$
\cite{:2008asa}. With the branch ratio for charm quarks to electrons
\cite{Eidelman:2004wy} and the correction for the geometrical acceptance,
the rapidity density of $c\bar{c}$ pairs can be estimated \cite{:2008asa}.
We use the PYTHIA event generator with the PHENIX acceptance to reproduce
the spectra from charm hadron contribution, see Fig. (\ref{fig:invariantmass-charm}(a)).
It can be seen that our results from PYTHIA (black-dashed line) are
consistent with those given by PHENIX (red-dotted line). We obtain
the cross section of $c\bar{c}$ pairs $\sigma_{c\bar{c}}=0.5$ mb.

\begin{figure}
\includegraphics[scale=0.4]{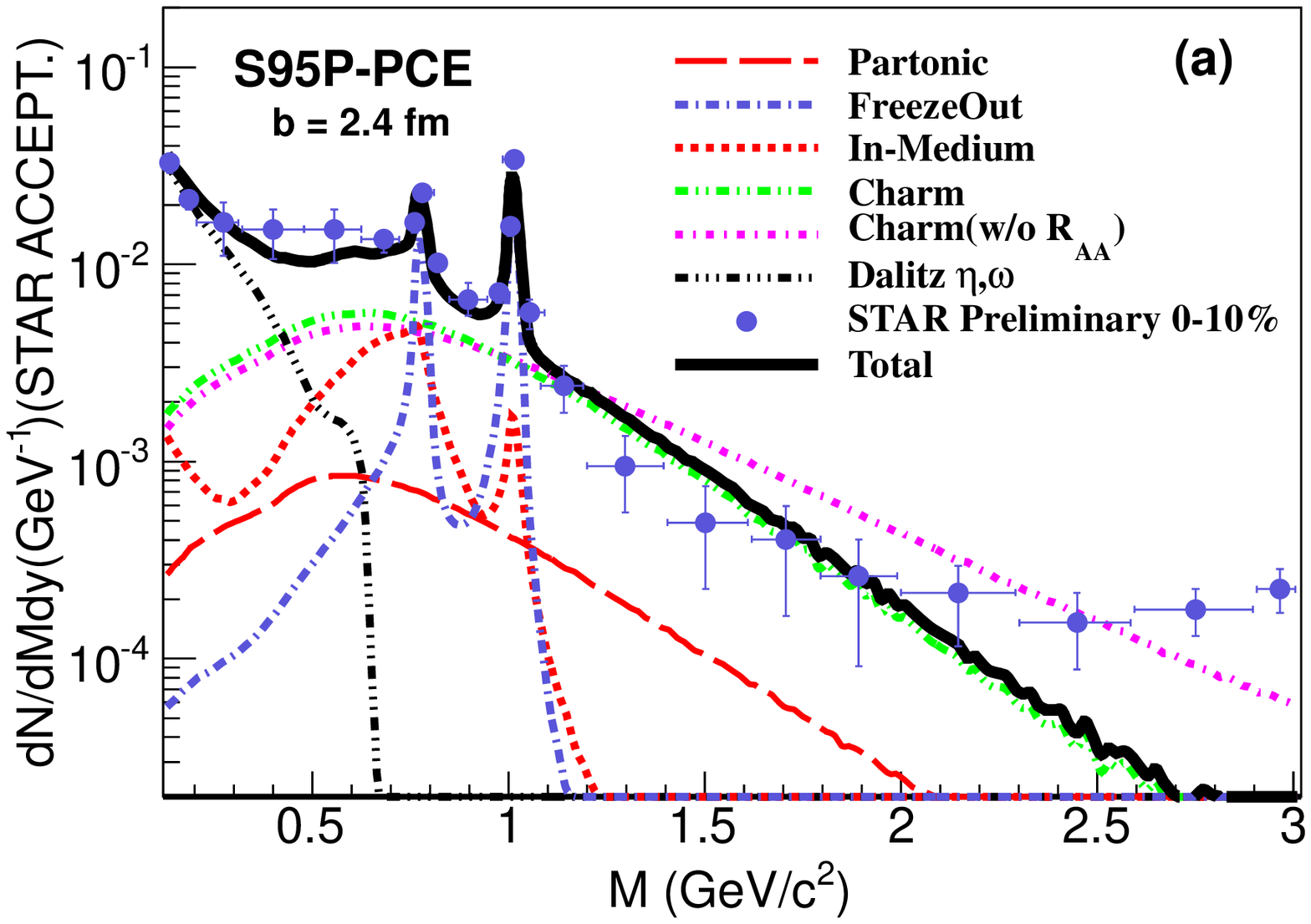} 

\includegraphics[scale=0.4]{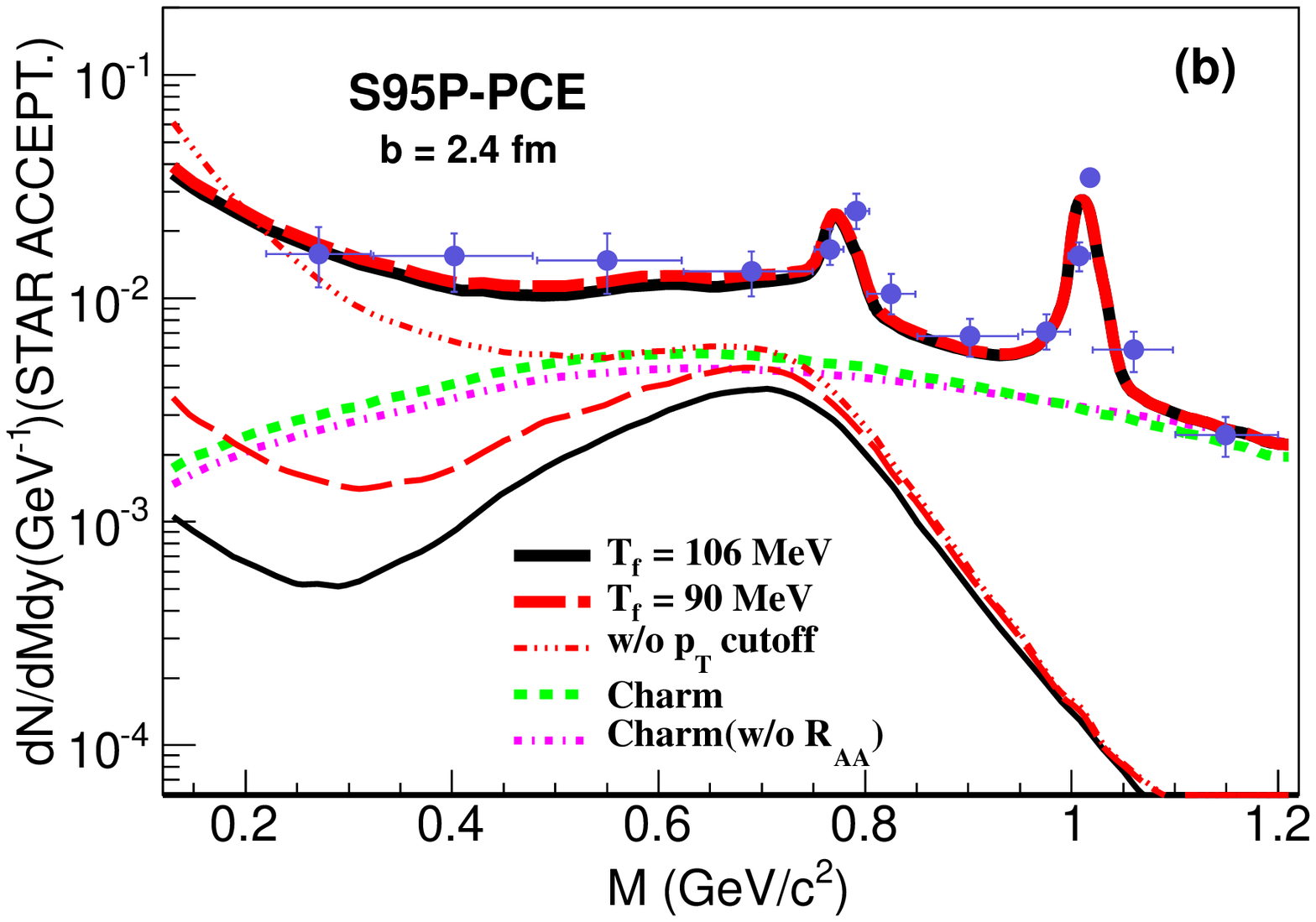}

\includegraphics[scale=0.4]{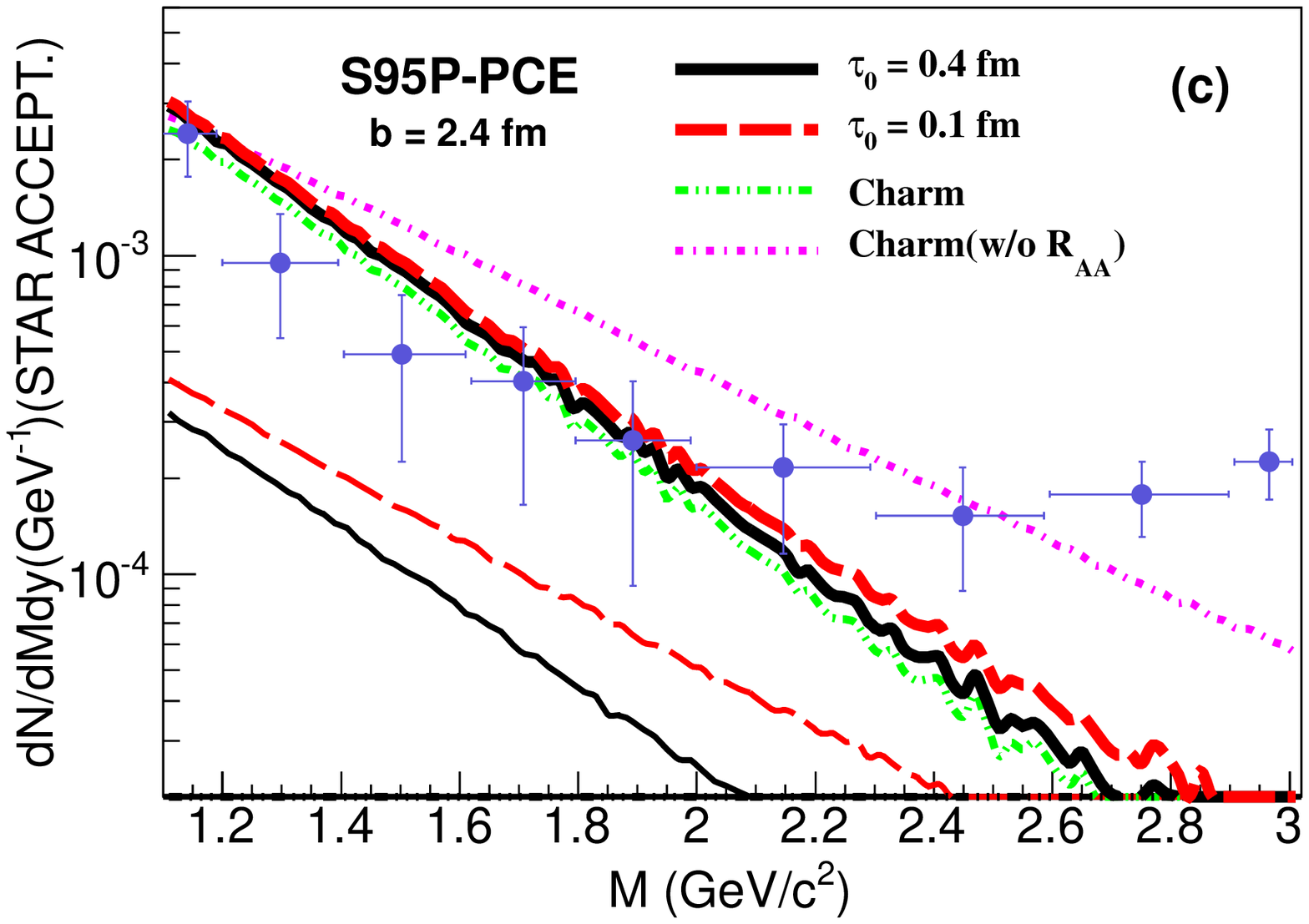}\caption{\label{fig:Comparison-with-RHIC}(Color online) The invariant mass
spectra and the comparison with STAR preliminary data \cite{Zhao:2011wa}
in most central (0-10\%) Au+Au collisions with the STAR acceptance.
The cocktail sums are in thick lines. See the text for detailed illustrations.
(a) The results in $M=[0.2,3]$ GeV. (b) The results in $M=[0.2,1.2]$
GeV. The thin-balck/thin-red-long-dashed line denotes the contribution
from the in-medium $\rho$ decays for $T_{f}=$106/90 MeV. The thin-red-dashed-dotted-dotted
line denotes the contribution from the in-medium $\rho$ decays without
the $p_{T}$ cutoff for $T_{f}=$90 MeV. (c) The results in $M=[1.1,3]$
GeV. The thin-balck/thin-red-long-dashed line denotes the QGP contribution
for $\tau_{0}=$0.4/0.1 fm. }

\end{figure}

For Au+Au collisions, we use the renormalized cross section in pp
collisions and scale it by the mean number of binary collisions. We
choose $N_{coll}=950$ for most central collisions. The charm quarks
are mostly generated in the pre-equilibrium stage. In medium the $p_{T}$
spectra of the charm quarks as well as the angular correlation of
the $c\bar{c}$ pairs could be modified due to its interaction with
the thermalized partrons. The medium modifications of heavy flavors
have been widely studied in, e.g., Ref. \cite{He:2011qa,Ghosh:2011bw}.\textcolor{red}{{}
}To include the medium modifications in a simple way, we parametrize
the nuclear modification factor of the single electron in the form,
\begin{equation}
R_{AA}^{e}(p_{T})=\mathrm{min}\left[1.0,\exp\left(a/p_{T}+b\right)\right],\label{eq:raae}\end{equation}
 where $a=1.23$ and $b=-1.51$, see Fig. (\ref{fig:invariantmass-charm}(b))
for the fitting function and experimental data. Note that we have
neglected the $\Upsilon$ contribution here. To get a realistic $p_{T}$
distribution for electrons, we use the original $p_{T}$ spectra obtained
by PYTHIA and multiply them with $R_{AA}^{e}(p_{T})$. Using the Monte
Carlo method, we sample the momentum spectra in accordance with the
resulting $p_{T}$ spectra for electrons and positrons respectively.
In each event we randomly choose from the sample the momenta of one
electron and one positron and combine them to a di-electron pair,
from which the invariant mass and total transverse momentum of the
pair can be determined. The modified invariant mass spectra by $R_{AA}^{e}(p_{T})$
are found to be narrower than without such a modification as shown
in Fig. (\ref{fig:Comparison-with-RHIC}(a)). 

With the STAR acceptance (transverse momentum $p_{T}>0.2$ GeV/c and
pseudorapidity $|\eta^{e}|<1$ for an individual electron, rapidity
$|y^{ee}|<1$ for a pair of electrons) and $p_{T}$ resolution, we
compute the di-electron spectra in most central collisions and compare
them with the STAR preliminary data of 0-10\% centrality, see Fig.
(\ref{fig:Comparison-with-RHIC}(a)). We also included the Dalitz
decay channels for $\eta$ \cite{Kroll:1955zu} and $\omega$: $\eta\rightarrow e^{+}e^{-}\gamma$
and $\omega\rightarrow e^{+}e^{-}\pi^{0}$. The $\eta$ contribution
can be easily deducted as a background in the experiment due to its
very long lifetime (about $1.5\times10^{5}$ fm/c), which leads to
its decay outside the freezeout scope. 

The cooktail sum including the in-medium $\rho$ mesons can roughly
reproduce the di-electron spectra in the LMR, see Fig. (\ref{fig:Comparison-with-RHIC}(a)).
In Fig. (\ref{fig:Comparison-with-RHIC}(b)), we show the total di-electron
spectra (thick lines), the contributions from the open charm (green-dashed
line) and the in-medium $\rho$ (thin lines) in the range $M\in[0,1.2]$
GeV/$c^{2}$. We found that the $\rho$ meson contributions (thin
black-solid line) are submerged under the open charm one. This indicates
that the charm backgrounds play an important role in the dilepton
spectra at the RHIC energy. As we discussed in Sect. (\ref{sec:Hydrodynamic})
we can tune $T_{f}$ to a lower value (e.g. 90 MeV in red-dashed line)
to increase the contribution from the in-medium $\rho$ mesons. The
cocktail sum (thick red-dashed line) with the in-medium $\rho$ contribution
(thin red-dashed line) for $T_{f}=90$ MeV is also shown. This seems
to give a better fit to the data. Though the lower $T_{f}$ gives
larger broadenings of the $\rho$ spectra and low mass enhancements,
the $\rho$ meson contribution is still smaller than the open charm
one. This is because: (1) The nuclear modification factor enhances
the charm contribution in the LMR; (2) Most of the low mass di-electrons
from the in-medium $\rho$ mesons have low $p_{T}$, which are beyond
the capability of the detectors and can not be measured. To support
the point (2), we calculate the in-medium $\rho$ meson contribution
incorporated by the STAR acceptance except the $p_{T}$ cutoff for
electrons and positrons. The result is shown in the thin red-dashed-dotted-dotted
line in Fig. (\ref{fig:Comparison-with-RHIC}(b)). We can see a strong
enhancement below the free $\rho$ mass. 

With the nuclear modification factor for charm hadrons, we can roughly
reproduce the di-electron spectra in the IMR, see Fig. (\ref{fig:Comparison-with-RHIC}(c)).
One can see that the thermal contributions from the QGP phase (thin
black-solid and red-dashed lines) are much smaller than the correlated
charm decays (blue-dashed and green-dashed-dotted-dotted lines). Now
we try to look at if it is possible to increase the QGP thermal contributions
in the IMR by the tuning parameters. We know that the thermal rate
from the QGP is proportional to $T^{4}A\tau$ where $A$ is the transverse
area \cite{Deng:2010pq}. To this end, in our model, we can tune the
equilibration time and entropy density (initial energy density) with
the constraint $s_{0}\tau_{0}$ = constant to keep the multiplicity
rapidity density unchanged. The dilepton emission rates do not change
much for different $\tau_{0}$. But for an earlier equilibration time,
e.g. $\tau_{0}=0.1$ fm (thin red-dashed line), the partonic contribution
is enhanced in the IMR, since the early equilibration time gives larger
space-time volume of high temperatures, whose di-electron emissions
mostly contribute to the IMR. But there is still a large gap between
the contributions from charm hadrons and from the QGP. In addition,
to lower the transition temperature will increase the space-time volume
of the QGP phase and then dilepton rates from thermal partons. But
this enhancement is almost in the LMR and will not significantly influence
the IMR. So it seems that it is very difficult to extract the thermal
sources from the backgrounds from charm hadron decays in the invariant
mass spectra alone. Additional observables such as $p_{T}$ spectra
and collective flows \cite{Renk:2006qr,Deng:2010pq} are also needed. 

\begin{figure}
\includegraphics[scale=0.4]{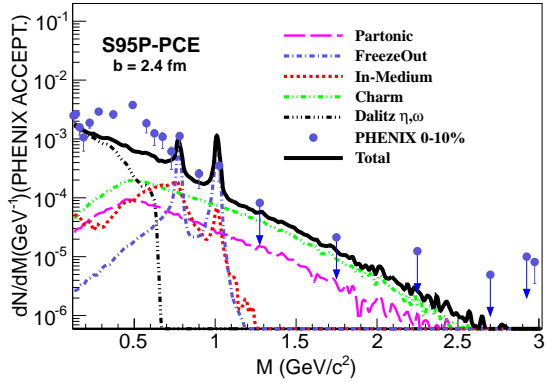}

\caption{(Color online) Comparison with PHENIX data \cite{Adare:2009qk} in
most central (0-10\%) Au+Au collisions with the PHENIX acceptance.\label{fig:PHENIX}}

\end{figure}

We show in Fig. (\ref{fig:PHENIX}) the results with the PHENIX acceptance
\cite{Adare:2009qk}. We see that the charm backgrounds still out-perform
the in-medium $\rho$. The acceptance geometry pushes the charm hadron
contributions toward the LMR. Using our cocktail sources, there is
still a large unexpected excess of di-electrons in the LMR as reported
by the PHENIX collaboration. 

\section{Summary and conclusion}

We investigate the di-electron low and intermediate mass spectra from
the vector and charm hadrons in most central heavy ion collisions
at ultra-relativistic energies. The space-time history of the fireball
is provided by a 2+1 dimension ideal hydrodynamic model, whose parameters
are fixed by fitting the transverse momentum spectra of long-life
hadrons, i.e., pions, kaons and protons. Two types of equations of
state are used. The medium effects of vector mesons from scatterings
of vector mesons by mesons and baryons in the medium are considered.
The di-electron emissions from in-medium vector meson decays can be
evaluated via the imaginary parts of the vector meson propagators
which are functions of space-time through the temperature. Due to
their longer lives than the time scale of the freezeout process, most
of the $\omega$ and $\phi$ mesons may decay at the thermal freezeout,
giving two sharp peaks in di-electron mass spectra. The contribution
from the charm hadrons is modeled by the PYTHIA simulation of the
proton-proton collisions and modified by the binary collision number
and the nuclear modification factor for electrons. 

The cocktail sum over all above sources and the partonic phase incorporated
with the acceptances of the STAR detector is compared to the STAR
preliminary data. The hadronic many body effective theory with a broadening
rho meson spectral function can describe the STAR di-electron data
in the LMR. With a parametrized nuclear modification factor for electrons
from charm hadron decays, we can roughly reproduce the di-electron
spectra in the IMR, though we still lack enough knowledge about open
charm decays in medium, such as the modification from the dynamical
correlation of $c\bar{c}$ pairs. 

In conclusion, we find: (1) The detector acceptance especially the
transverse momentum cutoff significantly suppresses the contribution
from the in-medium $\rho$ meson in the mass region below the $\rho$
mass; (2) With the current set of parameters and detector acceptances,
the backgrounds from charm hadrons dominate in the low and intermediate
mass regions. Therefore it is impossible to extract the thermal sources
of dileptons with the invariant mass spectra alone if the backgrounds
from charm hadrons are not removed. Other observables such as transverse
momenta and collective flows may provide additional tools to tag these
sources. Future STAR programs such as the Heavy Flavor Tracker \cite{Xu:2006zi}
and the Muon Telescope Detecctor are expected to improve the capability
of identifying the backgrounds from charm decays and extracting the
thermal souces. 

Acknowledgement: QW is supported in part by the National Natural Science
Foundation of China (NSFC) with grant No. 10735040. YFZ is supported
in part by the National Natural Science Foundation of China (NSFC)
with grant No. 10805046.

\bibliographystyle{apsrev}
\addcontentsline{toc}{section}{\refname}\bibliography{ref}

\end{document}